\documentclass[fleqn,usenatbib]{mnras}

\usepackage{newtxtext,newtxmath}
\usepackage{listings}
\usepackage{color}

\definecolor{dkgreen}{rgb}{0,0.3,0}
\definecolor{gray}{rgb}{0.5,0.5,0.5}
\definecolor{mauve}{rgb}{0.58,0,0.82}
\definecolor{golden}{rgb}{0.86,0.65,0.01}

\usepackage[T1]{fontenc}

\DeclareRobustCommand{\VAN}[3]{#2}
\let\VANthebibliography\thebibliography
\def\thebibliography{\DeclareRobustCommand{\VAN}[3]{##3}\VANthebibliography}

\usepackage{graphicx}	
\usepackage{amsmath}	
\usepackage{amssymb}	

\newcommand{\magrvs}{$\mathrm{G}_\mathrm{RVS}$}
\newcommand{\magg}{$\mathrm{G}$}
\newcommand{\magrp}{$\mathrm{G}_\mathrm{RP}$}
\newcommand{\col}{$\mathrm{G}-\mathrm{G}_\mathrm{RP}$}
\newcommand{\rvs}{GDR2$_\mathrm{RVS}$}
\newcommand{\all}{GDR2$_\mathrm{all}$}

\title[Gaia RVS selection function]{Characterising the Gaia Radial Velocity sample selection function in its native photometry}

\author[Jan Rybizki]{
Jan Rybizki,$^{1}$\thanks{E-mail: rybizki@mpia.de}
Hans-Walter Rix,$^{1}$
Markus Demleitner,$^{2}$
Coryn A.L. Bailer-Jones$^{1}$
\newauthor and William J. Cooper$^{3,4}$
\\
$^{1}$Max Planck Institute for Astronomy,
	K\"onigstuhl 17, D-69117 Heidelberg, Germany\\
$^{2}$Astronomisches Rechen-Institut, Zentrum f{\"u}r Astronomie der Universit{\"a}t Heidelberg, M{\"o}nchhofstrasse 12-14, D-69120 Heidelberg, Germany\\
$^{3}$School of Physics, Astronomy and Mathematics, University of Hertfordshire, College Lane, Hatfield AL10 9AB, UK\\
$^{4}$Istituto Nazionale di Astrofisica, Osservatorio Astrofisico di Torino, Strada Osservatorio 20, I-10025 Pino Torinese, Italy
}

\date{Accepted XXX. Received YYY; in original form ZZZ}

\pubyear{2020}

\begin{document}
\label{firstpage}
\pagerange{\pageref{firstpage}--\pageref{lastpage}}
\maketitle

\begin{abstract}
The Gaia DR2 radial velocity sample (\rvs), which provides six-dimensional phase-space information on 7.2 million stars, is of great value for inferring properties of the Milky Way. Yet a quantitative and accurate modelling of this sample is hindered without knowledge and inclusion of a well-characterized selection function.
Here we derive the selection function through estimates of the internal completeness, i.e. the ratio of \rvs\ sources compared to all Gaia DR2 sources (\all).
We show that this selection function or “completeness” depends on basic observables, in particular the apparent magnitude \magrvs\ and colour \col, but also on the surrounding source density and on sky position, where the completeness exhibits distinct small-scale structure.
We identify a region of magnitude and colour that has high completeness, providing an approximate but simple way of implementing the selection function. For a more rigorous and detailed description we provide python code to query our selection function, as well as tools and ADQL queries that produce custom selection functions with additional quality cuts.

\end{abstract}



\section{Introduction} \label{sec:intro}

Gaia DR2 contains median radial velocities and their uncertainties for 7 224 631 stars  \citep{2019A&A...622A.205K}. To fully exploit this data set, the astronomical community requires the \rvs\ selection function. A selection function quantifies the probability of an object entering a sample (here, the RVS sample) as a function of its observables, such as magnitudes, colours and position on sky. This is required for essentially all ensemble modelling of such data, as in such cases any model that predicts observables must first be multiplied with the selection function before a meaningful comparison to data is possible.
The characterisation of the \rvs\ selection function has been hampered as no \magrvs\ photometry, covering the far-red optical region of the RVS spectra, has been published. This is because of the Gaia spectroscopic pipeline not yet being fully calibrated  \citep{2018A&A...616A...6S}. 
Approximating \magrvs\ from \magg\ and \magrp\ magnitudes allows us to build the selection function: $\mathcal{S}\left(\mathrm{G}_\mathrm{RVS},\mathrm{G}-\mathrm{G}_\mathrm{RP},\left(\alpha,\delta\right)\right)$. \magrvs\ provides the brightness range over which the RVS instrument \citep{2018A&A...616A...5C} could collect enough signal over the Gaia DR2 time span and where the detector would not saturate. The \col\ colour is a proxy for the effective temperature of the star, for which reliable radial velocity determinations can be obtained from the measured spectral window. The sky position can enter the selection function via the source density, which varies dramatically across the sky, through the Gaia scanning law \citep{2020MNRAS.497.1826B,2020MNRAS.tmp.2380B}. Sky position also enters the selection function through a mix of both the RVS spectral window assignment on neighbouring RVS sources and (in the case of Gaia DR2) through a pre-selection of RVS sources based on other input catalogues \citet{2014A&A...570A..87S}.

 Our paper is structured as follows: in Section\,\ref{sec:native_photometry} we show how \magrvs\ can be derived. Since our empirical approach derives the {\it internal} completeness of the \rvs\ sample with respect to the \all\ sample, we show how this can be generalised to the {\it external} completeness, i.e. selection function in Section\,\ref{sec:internal_completness}.
 Section\,\ref{sec:CMD_completeness} looks at the completeness with colour and magnitude. In Section\,\ref{sec:source_density} we examine the completeness over the sky. Section\,\ref{sec:ill_defined} highlights correlations of the \rvs\ selection function with other parameters. We then explain the generation and usage of our \rvs\ selection function in Section\,\ref{sec:selection_function} and conclude with a summary in Section\,\ref{sec:summary}. 

\section{Using the native photometry of the RVS instrument}
\label{sec:native_photometry}
The Radial Velocity Spectrometer (RVS) is an integral field spectrograph \citep{2018A&A...616A...5C} that observes in the near-infrared at $\lambda=[845,872]$\,nm, which is redder than the mean wavelength of the \magg\ band ($\lambda=[330,1050]$\,nm) and also slightly redder than the mean wavelength of the \magrp\ band ($\lambda=[630,1050]$\,nm) \citep{2018A&A...616A...4E}.

\begin{figure}
	\includegraphics[width=\linewidth]{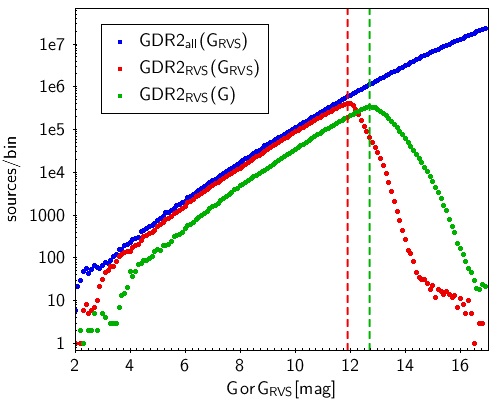}
	\caption{Magnitude distribution of \all\ (blue) and \rvs\ in different photometric bands (G in green and \magrvs\  in red). The mode of the \rvs\ distribution is indicated as a function of G and G$_\mathrm{RVS}$ magnitude at 12.7 and 11.9, respectively.}
	\label{fig:magdist}
\end{figure}

For Gaia DR2, the processing of radial velocities was limited to sources with \magrvs\,$<12$\,mag, but since no \magrvs\ magnitudes have been published this statement is ambiguous. To add to this confusion there are three different ways to determine \magrvs\ \citep{2018A&A...616A...6S,2019A&A...622A.205K}, which were used in different parts of the processing and affect the completeness of the \rvs\ sample:
\begin{itemize}
    \item The "on-board \magrvs" is derived on-board the satellite, is used for various on-board automated decisions, and is also transmitted to the ground\footnote{For sources with on-board \magrvs\,$>16.2\,$mag RVS windows are assigned, sources with on-board $<$ 7 get 2D windows assigned \citep{2018A&A...616A...6S}.}.
    \item The "external \magrvs" was calculated by \citet{2014A&A...570A..87S} from a collection of ground-based photometric catalogues. These \magrvs\ are part of the so-called Initial Gaia Source List (IGSL).
    \item The "internal \magrvs" is the magnitude derived by the Gaia spectroscopic pipeline using the flux recorded in the RVS spectra\footnote{Sources with internal \magrvs\,$>\,$14\,mag were not published in Gaia DR2.}.
\end{itemize}
Ideally we would like to use the "internal \magrvs" in this work but since it was not published we are instead forced to use yet another quantity which we call the \emph{approximated} \magrvs\ and was fit to the "internal \magrvs" after processing was finished \citet{2018A&A...616A...1G}.  
The approximated \magrvs\ is a function of \magg\ and \magrp\ as specified in Equations 2 \& 3 from \citet{2018A&A...616A...1G}, which gives G$_{\rm RVS}$-G$_{\rm RP}$ as a fourth order polynomial in G-G$_{\rm RP}$. These equations\footnote{can be inspected in the queries of Appendix\,\ref{sec:queries}} are valid within $0.1<\mathrm{G}-\mathrm{G}_\mathrm{RP}<1.7$ and approximate the true G$_\mathrm{RVS}$ to $\approx0.1$\,mag precision\footnote{We add 0.05\,mag at each limit to the colour range, i.e. $0.05<\mathrm{G}-\mathrm{G}_\mathrm{RP}<1.75$ in order for all bins to have the same width, i.e. 0.1\,mag.}. The advantage of switching to this approximated G$_\mathrm{RVS}$ value, which is closer to what RVS observes, is that it reduces the impact of colour variations on the selection function. This results in a sharper cut-off in the magnitude distribution which can be inspected in Figure\,\ref{fig:magdist}. For G$_\mathrm{RVS}$, 5.3M sources are brighter than the mode whereas for G this only holds for 4.5M sources.


To select the stars to be processed for Gaia DR2, the consortium used the external \magrvs, when available. For the 8\% of sources where it was not, the on-board \magrvs\ was used instead \citep{2018A&A...616A...6S}. 

The external \magrvs\ has been derived from multiple catalogues, with varying zero-points, and a multitude of photometric bands, resulting in multiple transformation formulas.
The sharp selection in external \magrvs\ translates into a more shallow one in the approximated \magrvs, though, as noted above, this is still sharper than when G is used. Throughout the paper we refer to the approximated \magrvs\ as \magrvs\ unless we add the prefixes from the 3 bullet points above\footnote{Only in the context of GeDR3mock \citep{2020PASP..132g4501R} is the \magrvs\ not approximated but calculated using an RVS passband.}. The tail of the \magrvs\ magnitude distribution comprises mostly stars with an external \magrvs\ brighter than 12\textsuperscript{th}\,mag, but which are fainter in the approximated \magrvs.

For the \magrvs\ distribution in Figure\,\ref{fig:magdist} we note that for the magnitude range 
\begin{equation}
\label{eq:mag}
 2.95 < \mathrm{G}_\mathrm{RVS} < 12.05  
\end{equation}
the \rvs\ sample follows the \all\ sample distribution quite well. Stars $<2.95\,$\magrvs\,mag do not usually enter the \rvs\ sample owing to saturation of the core of RVS spectra at approximately G\,$\sim 4$\,mag \citep{2019A&A...622A.205K}.

We limit our investigation to the colour range for which \magrvs\ can be approximated. From Table\,\ref{tab:starcounts_with_cuts} we deduce that we are only losing 0.2\,\% of \rvs\ sources because of the colour limit and most of them actually owing to a missing \magrp\ measurement.

\begin{table}
\caption{\rvs\ star counts within different cuts.}
    \centering
    \begin{tabular}{c|c}
     Within  & source number in millions \\
       \hline
       \rvs\ (all) & 7.225\\
       \magrvs\ approximation & 7.213\\
       high-completeness colour range (Eq.\,\ref{eq:col}) & 6.872\\
       high-completeness mag range (Eq.\,\ref{eq:mag})& 5.699\\
       high completeness CMD area (Eqs.\,\ref{eq:mag}\,\&\,\ref{eq:col}) & 5.443\\
\end{tabular}
\label{tab:starcounts_with_cuts}
\end{table}

\begin{figure}
	\includegraphics[width=\linewidth]{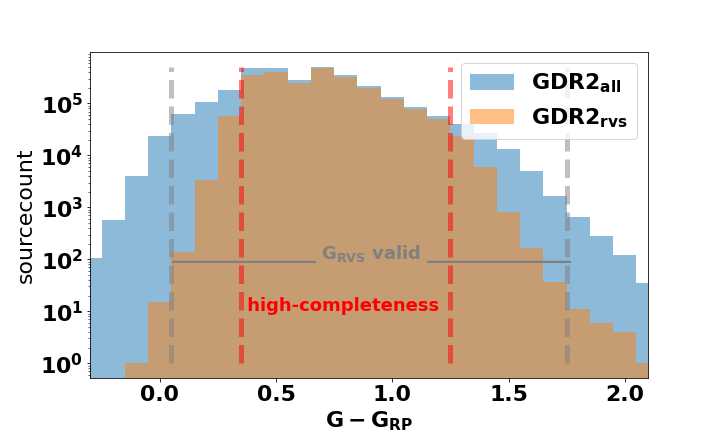}
	\caption{Colour distribution of \rvs\ and \all\ for $\mathrm{G}<12$\,mag, where faintness of the \magrvs\ magnitude should not play a role with respect to the internal completeness. The valid range of the G$_\mathrm{RVS}$ approximation is indicated with grey dashed lines. Outside the high-completeness colour range of Equation\,\ref{eq:col}, as highlighted by the red dashed lines, the vast majority of bright sources are not contained in \rvs.}
	\label{fig:color_range}
\end{figure}

In Figure\,\ref{fig:color_range} we compare the \all\ sources with \rvs\ sources where G\,$<12$\,mag in order to assess the \rvs\ sample completeness with colour. As we can see, only a minute fraction of \rvs\ sources reside outside the \magrvs\ approximation range (depicted with grey dashed lines). For a central colour range with
 \begin{equation}
 \label{eq:col}
  0.35 < \mathrm{G}-\mathrm{G}_\mathrm{RP} < 1.25.    
 \end{equation}
 (depicted with red dashed lines) we see that the \rvs\ sample is almost complete with respect to the \all\ sample. The reason for the decrease in completeness for very blue and red sources is that hot stars have strong Paschen lines affecting the determination of radial velocities from the Calcium triplet. Similarly, spectra of cool stars are dominated by TiO molecular bands which inhibit the determination of the pseudo-continuum at the shortest wavelengths \citep{2019A&A...622A.205K}. Therefore sources that needed RV templates\footnote{For 18\,\% of the \rvs\ sources, ground-based effective temperatures were used and the remaining were derived from the spectra themselves using only 28 templates \citep{2018A&A...616A...6S,2019A&A...622A.205K}. We did not find strong signatures in the selection function resulting from these two different ways of determining the template spectra.} with an effective temperature outside the range 3550 $<$ T$_{\rm eff}$ [K] $<$ 6900 were excluded in Gaia DR2. This translates into a relatively sharp colour dependence, which is blurred more at the red end owing to dust reddening as this extends the radial velocity determination to redder sources.
 
\begin{figure}
	\includegraphics[width=\linewidth]{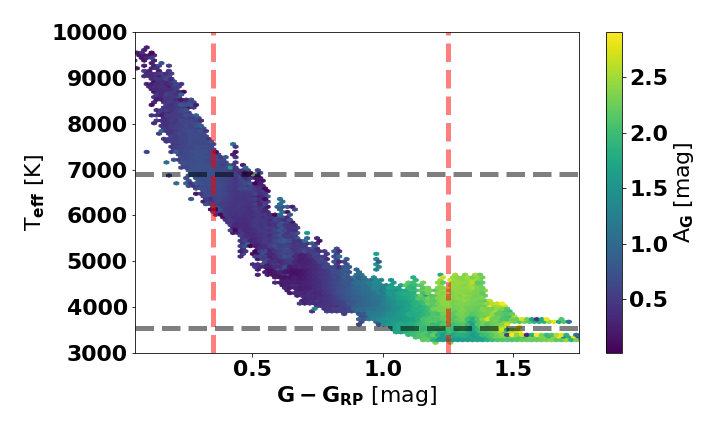}
	\caption{\col\,versus T$_\mathrm{eff}$ from GDR2, colour-coded by the mean extinction value for sources in the \rvs\ sample. Of the 7.2M sources, 4.8M have those estimates from GDR2 \citep{2018A&A...616A...8A}. Red dashed lines indicate the high-completeness colour from Equation\,\ref{eq:col}. The grey dashed lines show the T$_\mathrm{eff}$ limits for radial velocity determination.}
	\label{fig:teff_range}
\end{figure}

This can be seen in Figure\,\ref{fig:teff_range} where the \col\ versus effective temperatures are shown for the \rvs\ sample, colour-coded by the extinction estimate from GDR2 \citep{2018A&A...616A...8A}. Interestingly, the extinction estimates also increase for sources bluer than 0.35\,mag and therefore hotter than 6900\,K. Figure\,\ref{fig:teff_range} also illustrates that our defined high-completeness colour range from Equation\,\ref{eq:col} comes from the effective temperature range for radial velocity detection.

In the following, we will inspect in more detail the joint colour-magnitude dependence of the internal completeness of the \rvs\ sample. Ranges indicated by Equations\,\ref{eq:mag} and \ref{eq:col} will guide us where to expect nearly full completeness.

\section{Dependence on colour and magnitude}
\label{sec:CMD_completeness}
\begin{figure*}
	\includegraphics[width=\linewidth]{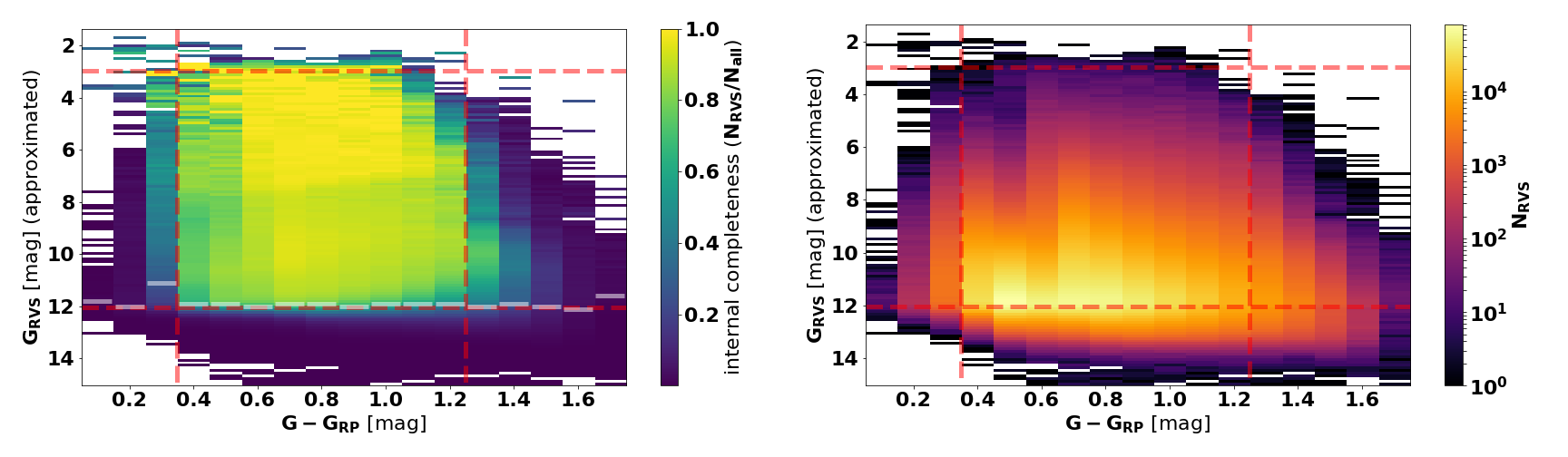}
	\caption{\rvs\ internal completeness in colour and magnitude in 0.1 x 0.1 mag binning. In the left panel, the colour-coding represents the ratio of \rvs\ versus \all\ , in the right panel the number of stars in \rvs. The white bars in the left panel indicates the modes of the magnitude distribution in each colour bin. The red dashed lines indicate the recommended 'high-completeness' ranges in colour and magnitude from Equations\,\ref{eq:mag}\,and\,\ref{eq:col}.}
	\label{fig:magcolorcut}
\end{figure*}

In order to obtain a better differentiated view of the \rvs\ sample completeness, we show the internal completeness, i.e. \rvs/\all\  source count, for colour-magnitude bins in the left panel of Figure\,\ref{fig:magcolorcut}. The red dashed lines show the colour magnitude ranges of high-completeness as specified in Equations\,\ref{eq:mag} \& \ref{eq:col}. For the \rvs\ sample we have 7.21M sources displayed (all sources within the \magrvs-approximation), of which 75\,\% are within the red dashed lines. For the \all\ sample, 6.57M sources are within the red dashed lines meaning that 83\,\% of the sources inside the red dashed lines have an RVS measurement. The colour bins that are adjacent but outside of the red dashed lines still have some higher fractions of completeness, but beyond these the completeness decreases fast. The only notable exception is the faint red corner at \magrvs$\,\sim 11$\,mag and \col$\,\sim 1.4$\,mag, which we attribute to dust-reddened sources with a well-defined Calcium triplet despite their red colour. The drop off for objects with \magrvs$\,>12$\,mag is quite sharp for all colour bins. There seems to be a decrease in completeness at \magrvs$\,\sim7$\,mag. This was the on-board \magrvs\ magnitude limit, where the Gaia pipeline switched from 2D to 1D window assignment \citep{2018A&A...616A...6S}. Those sources with \magrvs\ $>7$ are still relatively bright and produce spurious sources which also get 1D windows allocated by the sky-mapper. This then potentially produces window conflicts, which (because of the non-processing of truncated windows) leads to a lowered completeness in the range 7-9\,mag \magrvs, cf. Sec.3.2 \citet{2018A&A...616A...6S}, contrary to 2D windows which are processed even when blended. 

Curiously, the decrease in completeness at around 7\textsuperscript{th} \magrvs\ happens at a somewhat brighter \magrvs\ in the red than in the blue. This strongly indicates a colour-dependence of the on-board \magrvs\ estimate. The lowered completeness vanishes for fainter sources as these produce less spurious sources. In crowded regions of the sky (especially where the number of visits is still low, i.e. towards the Galactic center and the anti-center\footnote{This can be well seen when inspecting the HEALpix CMDs of \url{https://www2.mpia-hd.mpg.de/homes/rybizki/Jan\%20Rybizki\%20-\%20Homepage_files/rvs_selection_visualisation_intermap.html} switching between regions of high and low median transits as in Figure\,9 of \citet{2019A&A...622A.205K}.}) the effect of lowered completeness at the beginning of the 1D window assignment is worst, cf. left panel of Figure\,\ref{fig:skydensity}.

In the right panel of Figure\,\ref{fig:magcolorcut} the source density of the \rvs\ sample per colour-magnitude bin is shown. Here we see that in absolute numbers there are still a significant number of \rvs\ sources that have \magrvs\,$>12$\,mag or \col$\,>1.25$\,mag.

\section{Dependence on source density and sky position}
\label{sec:source_density}
Here we investigate the dependence of the selection function with respect to the position on the sky $\mathcal{S}\left(\left(\alpha,\delta\right)\right)$. As we will see in Section\,\ref{sec:density}, this has imprints of the global source density, as well as the close RVS pairs which compete in the spectral window allocation. Furthermore the dependence on input catalogues is still visible in Gaia DR2 as we will see in Section\,\ref{sec:sky_completeness}.
In the generation of our selection function, the sky position enters through HEALpix bins while the source density is only accounted for indirectly by lowering the HEALpix level until enough sources are in a respective selection function bin. See Section\,\ref{sec:selection_function} for details.
\subsection{Source density}
\label{sec:density}
\subsubsection{Global}
\begin{figure}
	\includegraphics[width=\linewidth]{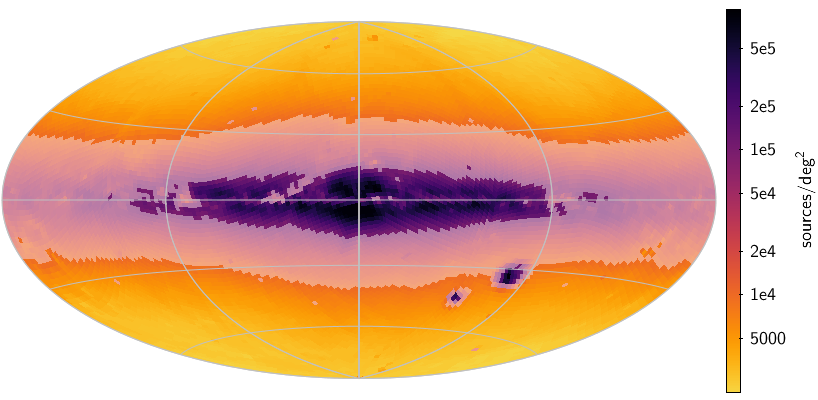}
	\caption{Aitoff projection of the sky density in Galactic coordinates. The Galactic Center is in the middle, with longitude increasing to the left at HEALpix level 5. This Figure shows how we split the \rvs\ sample into three source density regimes: low density (\rvs\  1.4M; \all\  100.3M), intermediate density (\rvs\  3.5M; \all\  540.0M) and high density (\rvs\  2.4M; \all\ 1,052.6M). The intermediate density area is shown with pale transparency; the high density part is towards the Galactic plane and the Magellanic clouds. The low density parts are at higher Galactic latitudes.}
	\label{fig:sky_partition}
\end{figure}

To the first order, the completeness is driven by projected stellar density (we neglect the sources from the second telescope in our work, but discuss its contribution in Appendix\,\ref{sec:FOV2}) which results in truncated RVS windows that have not been processed in Gaia DR2 \citep{2019A&A...622A.205K}. Also the processing limit of the RVS instrument is around 36k sources per degree$^2$ \citep{2018A&A...616A...5C} (though in the \rvs\ sample the highest density at HEALpix level 6 is 2681 sources/degree$^2$). We therefore split our sample into high- ($>$ 100k sources / degree$^2$), intermediate- and low- ($<$ 10k sources / degree$^2$) density regions on the sky (HEALpix level 5). See Figure\,\ref{fig:sky_partition} for their distribution.

\begin{figure*}
	\includegraphics[width=\linewidth]{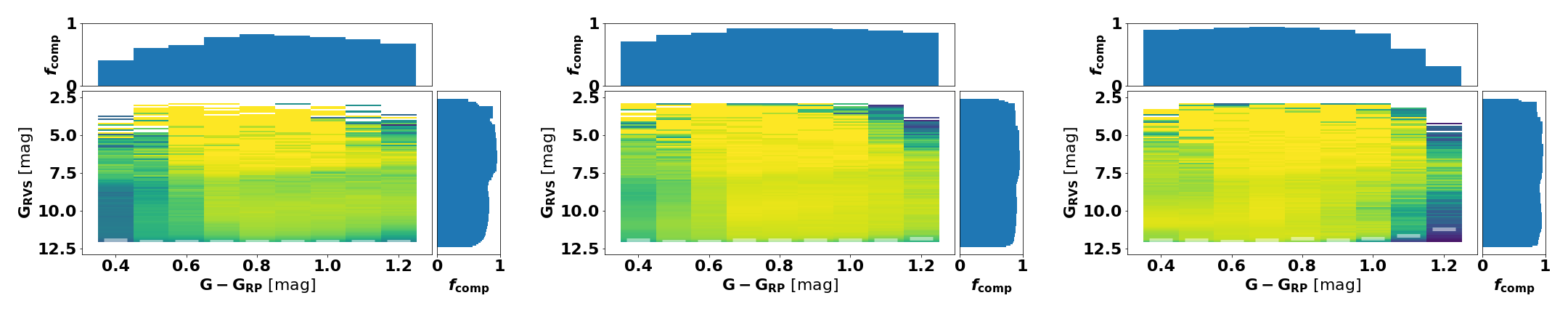}
	\caption{Same as left panel of Figure\,\ref{fig:magcolorcut} (using the same colour-scale) but in density subsets and only within the recommended high-completeness colour and magnitude ranges, see Eqs.\,\ref{eq:mag} \& \ref{eq:col}. From left to right are the high-, intermediate- and low-density sample with sources / degree$^2$ $>$ 100k, 10k $<$ and $<$ 100k, and $<$ 10k respectively. The white bars show the mode of the magnitude distribution per colour bin. \emph{Fcomp} represents the fractional completeness marginalised over colour or magnitude. The number of \rvs\ sources (\all\ sources) for these Figures are 1.55M (2.12M), 2.72M (3.14M) and 1.18M (1.30M), from left to right (the HEALpix fractions are 9\,\%, 39\,\% and 52\,\%). The partition over the sky is shown in Figure\,\ref{fig:sky_partition}.}
	\label{fig:skydensity}
\end{figure*}

In Figure\,\ref{fig:skydensity} the resulting CMDs for the high-, intermediate- and low-density sample are shown from left to right. As expected the overall completeness increases with decreasing density from 73\,\% over 87\,\% to 91\,\%, respectively. For the high-density sample (left panel) the drop off in completeness for sources with \magrvs$\,<7$\,mag is more pronounced and the faint blue corner has particularly low completeness, owing to the correlation between high-density and dust-reddened regions. For the intermediate density sample (middle panel) only the faint blue corner has a somewhat lowered completeness. For the low-density sample (right panel) the faint red end has lowered completeness, which is owing to the correlation between low-density and unreddened sky areas.

\subsubsection{Close pairs}
There are fundamental limitations of the Gaia satellite for the detection of nearby sources (sky separation), the so called 'contrast sensitivity', which is a function of projected distance and magnitude difference \citep{2015A&A...576A..74D}. For \all\ sources with a G magnitude and a sky position measurement this has been characterised in \citet{2019A&A...621A..86B}, which is at approximately 0.4\arcsec\ for equal brightness sources. When requiring a colour measurement this distance increases to approximately 2.0\arcsec\ \citep[Fig.9][]{2018A&A...616A..17A}\footnote{except for closer equal brightness binaries which still entered the catalogue}, because of truncated windows, which have not been included in Gaia DR2 photometry \citep{2018A&A...616A...3R}.

For the \rvs\ sample there is an additional factor at the processing stage owing to its extremely elongated windows\footnote{RVS spectral windows are 10 across-scan pixels times $\sim$1.3k along-scan pixels wide, which corresponds to 1.77\arcsec\ times $\sim$75\arcsec\ \citep{2012AN....333..453P}. This also means that a single RVS spectrum effectively takes up 135.4 arcseconds squared on the sky. If tightly packed this would allow for approximately 100k sources per degree$^2$ which is a factor of 3 higher than the instrument limit. In \rvs\ sample the highest density per degree$^2$ is 2681. If isotropically distributed the mean distance of neighbouring sources in the highest density HEALpix would be approximately 55\arcsec.} and relatively small source densities. The latter will change with future data releases, when going from a GRVS limit of 12 mag in GDR2, to 14 mag in GDR3, and perhaps as faint as 16 mag in GDR4. The maximum source densities per degree$^2$ will increase from 2.7k in GDR2 to (a theoretical maximum of) 50k in GDR3 or 300k in GDR4 (though the instrument limit is reached at 36k). Sources in denser areas are lost because deblending was not yet activated and truncated windows which were not rectangular were not processed \citep{2018A&A...616A...6S}. This does not apply to 2D windows, which were assigned to sources with \magrvs\ $<$ 7\,mag.

In Figure\,\ref{fig:contrast_sensitivity}\footnote{The query for the close pair data can be inspected in Appendix\,\ref{sec:queries} where we also provide a link to the data which includes all pairs between \rvs\ sources and \all\ sources.} we show log density plots of the sky separation versus \magrvs\ difference for \rvs\ sources. We show this for high density (left panel) and low density sky areas (right panel) as defined by Figure\,\ref{fig:sky_partition}. Owing to the low-number statistics and relatively low \magrvs\ range of the sample the contrast sensitivity is not well sampled. In high density areas chance alignments, which increase with distance squared, dominate the close pairs. For low density areas these play less of a role and true binaries in very close vicinity (over-density up to 5\arcsec) are more abundant. The minimum separation of two \rvs\ sources\footnote{this does not apply for sources with 2D windows} is 5 across-scan pixel \citep{2018A&A...616A...6S} corresponding to 0.85\arcsec\ \citep{2012AN....333..453P}). The very close true binaries (1-2\arcsec) are less well sampled in the high density regions indicative of source loss because of window truncation.

We do not attempt to include the effect of close pairs into our selection function, still from Figure\,\ref{fig:contrast_sensitivity}, one can approximate one if needed for binary treatment, e.g. : 
\begin{equation}
\mathcal{S}_\mathrm{close\ pairs} = \begin{cases}
    0     & \text{ if } \Delta\mathrm{G}_\mathrm{RVS} > 0.5\,\mathrm{mag}\ \wedge \ \mathrm{dist} < 2\arcsec \\
    1  & \text{ else }  \\
\end{cases}.
\end{equation}

\begin{figure*}
\includegraphics[width=\linewidth]{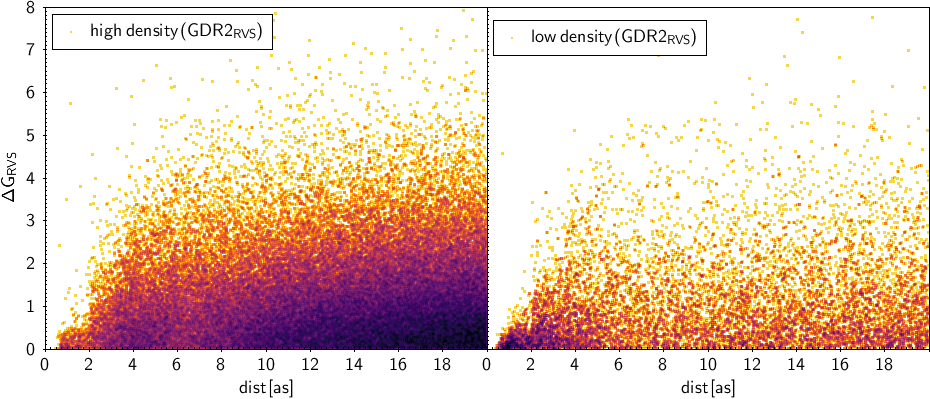}
 	\caption{Density plots (colour scale is in log and not the same for both panels) of the distance versus \magrvs\ difference for \rvs\ sources. The left panel shows high density areas with more than 100k sources per degree$^2$. For the right panel these close pairs are shown for low density areas on the sky (less than 10k sources per degree$^2$). The number of pairs for each of those subsets (not all are necessarily depicted) are 86k and 10k, respectively.}
	\label{fig:contrast_sensitivity}
\end{figure*}


\subsection{Sky position}
\label{sec:sky_completeness}
It is important to recall for the following subsections that the external \magrvs\ , on which the magnitude limit of 12 for most of the sources was applied comes from the IGSL3 \citep{2014A&A...570A..87S}. This used several catalogues and transformation formula in generating an \magrvs\ estimate, which we defined as external \magrvs\ in this paper. IGSL3 has 15M sources that have \magrvs\,$<12$\,mag. The \magrvs\ determination was prioritised in order of the following input catalogue list (percentage of IGSL sources with external \magrvs\,$<12$\,mag in brackets): SDSS  \citep{2002AJ....124.1810S} (4), TYCHO2 \citep{2000A&A...355L..27H} (16), GSC23 \citep{2008AJ....136..735L} (80) and negligible fractions from other input catalogues.  

\begin{figure*}
	\includegraphics[width=\linewidth]{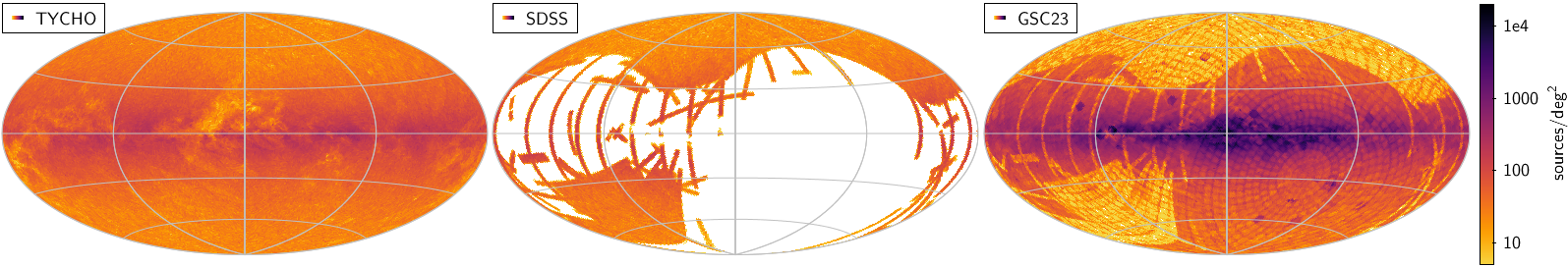}
	\caption{Source densities from different catalogues (Aitoff projection as in Figure\,\ref{fig:sky_partition} at HEALpix level 6). From left to right TYCHO, SDSS and GSC23 (Guide Star Catalog, see text) footprint in the IGSL (Gaia Input Catalog, see text) for sources with \magrvs\,$<12$\,mag. Colour-scale is fixed to cover the range of all three panels.}
	\label{fig:igsl_footprint}
\end{figure*}

In Figure\,\ref{fig:igsl_footprint} we can see that the priorities in choosing from different catalogues results in structure that we will recognise in the \rvs\ completeness function. The TYCHO set on the left has the highest priority and appears well-behaved. The SDSS footprint in the middle panel does not cover the whole sky. Areas and magnitude ranges that are left out are filled with the GSC23 data from the right hand panel.

\begin{figure*}
	\includegraphics[width=\linewidth]{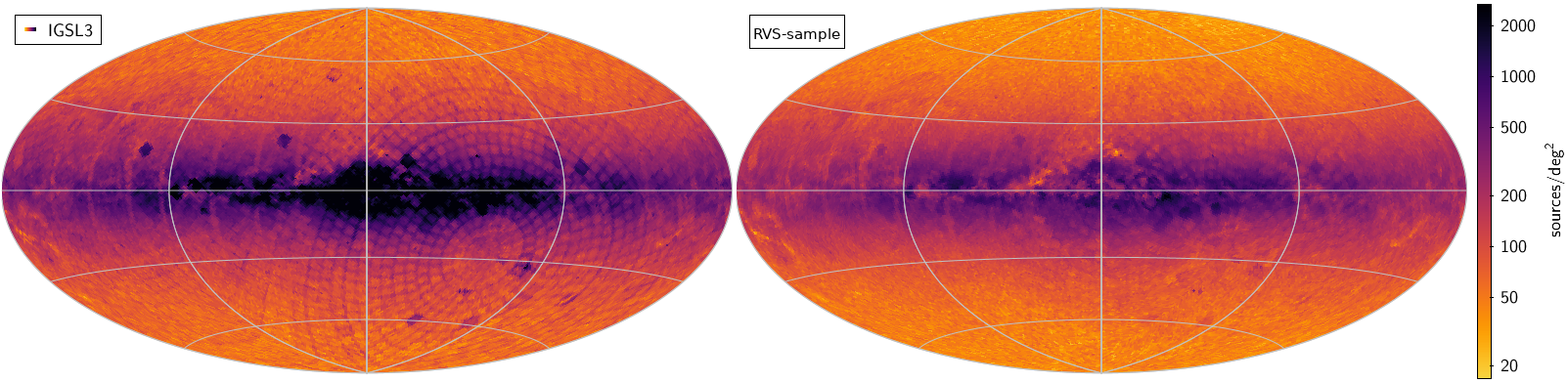}
	\caption{Source densities for IGSL3 with \magrvs\,$<12$\,mag on the left and the \rvs\ sample on the right (Aitoff projection same as in Fig\,\ref{fig:sky_partition} at HEALpix level 6). The colour-bar range is fixed to the \rvs\ sample, which has limits 17 and 2681 sources/degree$^2$; the colour bar saturates for the IGSL3 catalogue.}
	\label{fig:igsl_rvs}
\end{figure*}

This results in the overall IGSL footprint seen in the left panel of Figure\,\ref{fig:igsl_rvs}. For comparison we show the \rvs\ sample density on the right. Differences are mainly a result of overlapping windows in dense areas (and therefore lost sources in the \rvs\ sample on the right) and for 8\,\% of \rvs\ sources also the on-board \magrvs\ has been used instead of the external \magrvs. But still patterns from the GSC23 and SDSS footprint can be recognised in the \rvs\ sample.

\subsubsection{Magnitude dependence of spatial completeness}
\begin{figure*}
	\includegraphics[width=\linewidth]{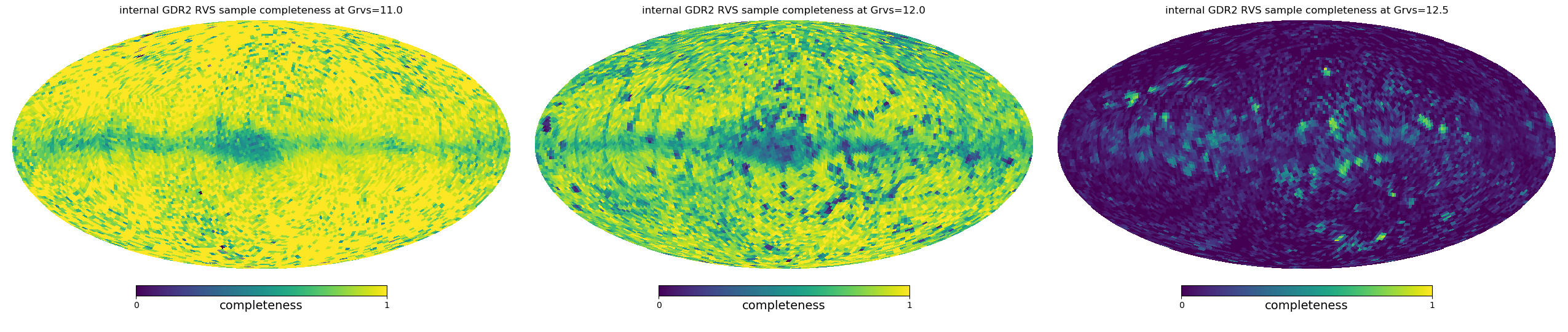}
	\caption{Internal completeness of the \rvs\ sample over the entire sky (as a Mollweide projection as in Fig.\,\ref{fig:sky_partition} at HEALpix level 5) at different G$_\mathrm{RVS}$\,magnitudes, with yellow being 100\,\% complete and dark blue being 0\,\% complete.}
	\label{fig:magsky}
\end{figure*}

In Figure\,\ref{fig:magsky} we show the completeness over sky with $\mathrm{G}_\mathrm{RVS}$ magnitude when only using sources within the high-completeness colour ranges of Equation\,\ref{eq:col}. A video version can be downloaded from here\footnote{\url{https://keeper.mpdl.mpg.de/f/db4ca9ee9bc34513bed1/}}.

At G$_\mathrm{RVS}=11$\,mag (left panel) the completeness shows little spatial structure, with only slightly lower completeness towards the Galactic plane. What can be already recognized is the SDSS footprint (lower completeness stripes perpendicular to the Galactic plane) that can also be seen in the left hand panel of Figure\,\ref{fig:igsl_rvs}, which shows the source density of the IGSL catalogue with external \magrvs\,$<12$\,mag. Since the SDSS transformation was preferentially used when calculating the sources in these stripes (cf. Figure\,\ref{fig:igsl_footprint} middle panel), it likely means that the SDSS transformation results in a fainter external \magrvs\ magnitude compared to the GSC23 transformation. This is consistent with Figure 16 of \citet{2018A&A...616A...4E} where the G magnitude estimate from SDSS data varies over the sky and seems to be especially different in parts of the stripes.

At 12\textsuperscript{th} G$_\mathrm{RVS}$\,mag (middle panel of Figure\,\ref{fig:magsky}), small patches of lower completeness start to emerge. These areas are point-like and distributed over the entire sky. It could be that these spots are correlated with brighter zero points in the respective GSC23 plates.

At G$_\mathrm{RVS}=12.5$\,mag there are still a few areas of high completeness visible. The circular glass-canopy pattern that can be seen is a result of the derivation of most of the external \magrvs\ magnitudes from the GSC23 Bj and Rf magnitudes \citep{2008AJ....136..735L}. The magnitude zero-points of the plates seem to be a little fainter at the edges (possibly where they overlap each other, private communication David Katz). The canopy pattern delineates the borders of the plates. It is also clearly visible in both panels of Figure\,\ref{fig:igsl_rvs}. There also seem to be plates for which the total area has a fainter zero-point. 

The star counts for the \rvs\ (\all) in the 0.1\,mag bins in Figure\,\ref{fig:magsky} are 202k (234k), 391k (569k) and 114k (863k) from left to right.

\subsubsection{Colour dependence of spatial completeness}
\begin{figure*}
	\includegraphics[width=\linewidth]{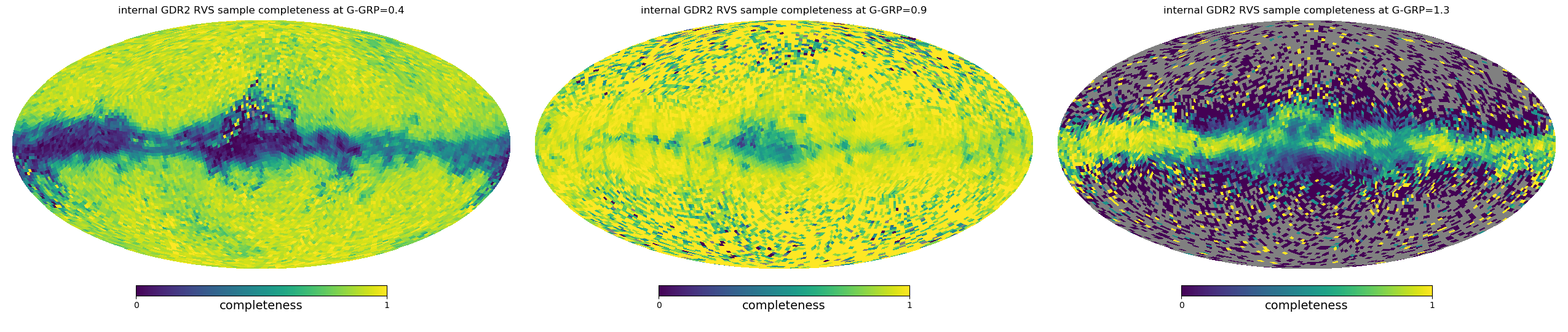}
	\caption{Internal completeness of the \rvs\ sample over the entire sky (as a Mollweide projection as in Fig.\,\ref{fig:sky_partition} at HEALpix level 5) as a function of colour, $\mathrm{G}-\mathrm{G}_\mathrm{RP}$ marginalized over the magnitude limits, $2.95\,\mathrm{mag}<\,$\magrvs$\,<12.05$\,mag as specified in Equation\,\ref{eq:mag}.}
	\label{fig:colorsky}
\end{figure*}

In Figure\,\ref{fig:colorsky} we show the completeness over sky with $\mathrm{G}-\mathrm{G}_\mathrm{RP}$ when only using sources within the high-completeness magnitude range of Equation\,\ref{eq:mag}. A video version can be downloaded from here\footnote{\url{https://keeper.mpdl.mpg.de/f/edbe4cc51d544b738ab5/}}.

In the left hand panel of Figure\,\ref{fig:colorsky} where the completeness for the colour bin at \col\,$=0.4$\,mag is shown, we can see notably lower completeness wherever dust reddening is in place. This is owing to hot stars which have no treatment from the RVS pipeline getting reddened to colours that would usually (in the absence of dust-reddening) result in a radial velocity measurement; this can be seen by the good completeness out of plane. A second order effect can be be witnessed in lower completeness areas that can also be seen but are even less pronounced in the left hand panel of Figure\,\ref{fig:magsky}, for instance, the stripe from the Galactic south pole left up to approximately l\,$=90$ and $b=0$. This feature and similarly lower completeness areas in the top right, Galactic anti-center, etc. are because of the scanning law and the necessity of at least 2 transits for a radial velocity measurement\footnote{The \rvs\ sample has a hard lower limit of 2 transits, a median of 7 and a maximum of 201.}, cf. Figure\,9 of \citet{2019A&A...622A.205K} or Figure\,5 panel (c) and (f) of \citet{2020MNRAS.tmp.2380B}.

The middle panel of Figure\,\ref{fig:colorsky} is similar to left panel of Figure\,\ref{fig:magsky} with the SDSS stripes of lower completeness and a lower completeness in the Bulge because of blends; interestingly enough however, the Galactic plane overall suffers less from incompleteness. When going redder in the right panel the completeness towards the plane improves the most. The reason for this counter-intuitive effect is because sources in the good temperature regime (\(3550<\mathrm{T}_\mathrm{teff}\,[K]<6900)\) for Ca triplet radial velocity determination fall into this colour bin (i.e. \col\,$=1.3$\,mag) only when dust-reddened and therefore in the Galactic plane.

The star counts for the \rvs\ (\all) in the 0.1\,mag bins are 533k (746k), 626k (731k) and 115k (205k) from left to right.

\section{Correlation with other quantities}
\label{sec:ill_defined}
We want to assess how well the \rvs\ sample is representative of the \all\ sample in terms of Gaia derived quantities as a function of both magnitude and colour. If those Gaia derived quantities were similar, one could apply a simple completeness correction by up-sampling the \rvs\ sources to the numbers of \all\ sources in their respective CMD bin. 

\begin{figure*}
	\includegraphics[width=\linewidth]{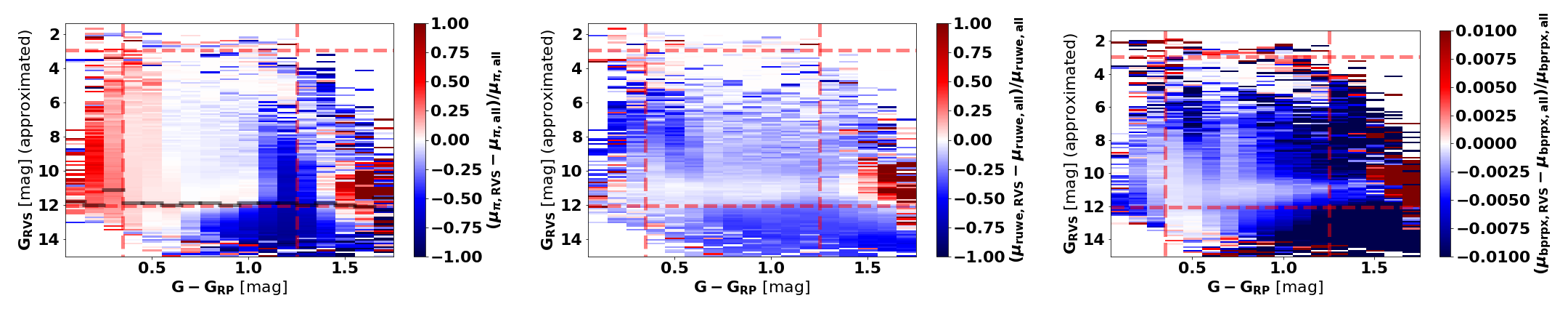}
	\caption{G$_\mathrm{RVS}$ magnitude vs \magg$\,-\,$\magrp\ colour, colour-coded by the fractional difference of mean ($\mu$) parameter values for the \rvs\ sample vs the \all\ sample in the respective colour-magnitude bin.  From left to right the parameters are \texttt{parallax}, \texttt{ruwe} and \texttt{phot\_bp\_rp\_excess\_factor}, note the different colour-scale for the right panel. The grey bars in the left panel show the mode of the magnitude distribution per colour bin.}
	\label{fig:bias_in_rvs}
\end{figure*}

In Figure\,\ref{fig:bias_in_rvs} we colour-code the fractional difference of the mean \texttt{parallax}, \texttt{ruwe}\footnote{The Renormalised Unit Weight Error (RUWE) is expected to be around 1.0 for sources where the single-star model provides a good fit to the astrometric observations. A value significantly greater than 1.0 (e.g. $>$\,1.4) could indicate that the source is non-single or otherwise problematic for the astrometric solution.} and \texttt{phot\_bp\_rp\_excess\_factor}\footnote{The BP/RP excess factor is the sum of the integrated BP and RP fluxes divided by the flux in the G band (BP and RP are dispersed and therefore more prone to light from nearby sources than G). This excess is believed to be caused by background and contamination issues affecting the BP and RP data. Therefore a large value of this factor for a given source indicates systematic errors in the BP and RP photometry.} (from left to right) between the \rvs\ sample and the \all\ sample on the CMD. Comparing to the left hand panel of Figure\,\ref{fig:magcolorcut} we see that lower values of internal completeness generally coincide with differences of the \rvs\ sample to the \all\ sample for the shown parameters, and the patterns generally differ between the parameters. For the parallaxes it seems that sources with \col\,$<0.35$ that are in the \rvs\ sample have generally higher parallaxes compared to the \all\ sample whereas for sources with \col$\,>1.05$ the opposite applies. The reason for this is that blue sources which are dust-reddened are usually further away whilst simultaneously being too hot for a radial velocity determination (cf. the left hand panel of Figure\,\ref{fig:colorsky} where for blue sources a low completeness in the Galactic plane areas can be seen). Therefore, blue sources in \rvs\ are generally closer to us than the blue sources in \all. Vice versa, the red sources ($1.05<$\,\col$\,<1.35$) in the \rvs\ sample are dust-reddened bluer stars for which a RVS measurement is possible (cf. right panel of Figure\,\ref{fig:colorsky}). Those are usually further away than unreddened sources of the same colour for which molecular lines in the spectrum prohibit the RVS measurement. This of course also means that the proper motions are different between those two samples, potentially biasing kinematic selections.

For the \texttt{ruwe} and the \texttt{phot\_bp\_rp\_excess\_factor} it seems that generally the \rvs\ sample has better quality sources than the overall \all\ sample.
There is one exception to this rule at the red end (outside of
the high-completeness zone) for \magrvs$\,\approx 11$\,mag where quality indicators are better than for the \all\ sample. However, this only applies to a tiny fraction of the \rvs\ sources.

Therefore we recommend to not use bins of the completeness function with completeness below some threshold, e.g. 90\,\%. If a larger sample is needed because of a weak signal we caution that the \rvs\ sample properties might not be representative of the \all sample.

\subsection{RVS selection in GeDR3mock with parallax bias}
In order to assess the impact of the parallax bias on the spatial distribution of the \rvs\ sample we applied the completeness function [HEALpix level 5, 0.1 \magrvs\ bins, 0.1 \col\  bins] to the GeDR3mock catalogue \citep{2020PASP..132g4501R}. We randomly chose 1000 subsets in each bin and took the subset that had the mean parallax closest to the mean of the respective:
\begin{itemize}
    \item (a) \rvs\ bin
    \item (b) \all\ bin.
\end{itemize}
And also (c) a truly random subset, which we do not consider here, but can, together with (a) and (b), be retrieved from here\footnote{\url{http://dc.g-vo.org/browse/gedr3mock/q}}.

\begin{figure*}
\includegraphics[width=\linewidth]{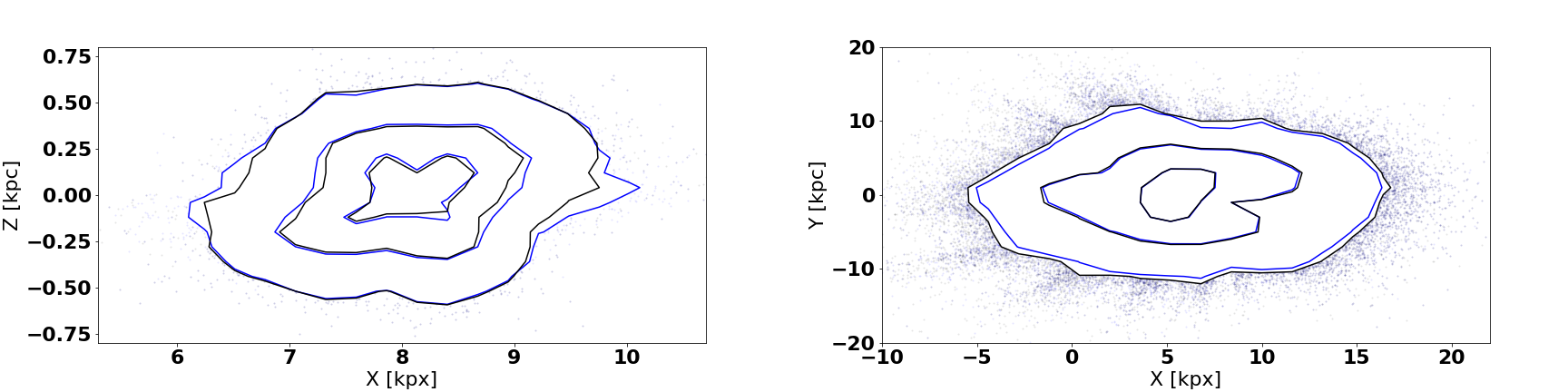}
 	\caption{Spatial distribution of a colour subset of the GeDR3mock RVS sample with parallaxes that are similar to the \all\ sample mean (per HEALpix, mag, colour bin) in blue and with parallaxes that are similar to the \rvs\ sample mean in black. The left panel shows 57k stars that are in the colour-range: $0.15\,\mathrm{mag}<\,$\col\,$<0.35$\,mag on the right panel the 743k stars with a colour-range of $1.05\,\mathrm{mag}<\,$\col\,$<1.35$\,mag are shown. The contour lines represent the (1,2,3)-$\sigma$ levels encompassing (39.3, 86.5, 98.9)\ \% of the stellar density distribution projected onto the respective Galactic plane, i.e. XZ on the left and XY on the right.}
	\label{fig:spatial_distribution_rvs}
\end{figure*}

In Figure\,\ref{fig:spatial_distribution_rvs} we display the spatial distribution of sample (a) black lines versus sample (b) in blue lines in. The left panel shows the Galactic XZ projection for the blue sources ($0.15\,\mathrm{mag}<\,$\col\,$<0.35$\,mag) where we see that in the Galactic plane the \rvs\ sample (a) does not probe as deep as the \all\ sample (b) with about 0.3\,kpc difference at the outer (3-$\sigma$) density contour. Similarly, the right panel shows the Galactic XY projection for the red sources ($1.05\,\mathrm{mag}<\,$\col\,$<1.35$\,mag) where we see that the \rvs\ sample probes approximately 1\,kpc further than the \all\  sample on the far side of the Galaxy at the outer (3-$\sigma$) density contour.

The difference is less pronounced than it probably is with the real Gaia data because we only chose from random subsets of sources from GeDR3mock; this choice does not perfectly represent the \all\ parallax distribution. Nevertheless, it should be beneficial to investigate those subsets and see how other selection effects can bias a sample, e.g. cuts on fractional parallax uncertainty.

\section{From internal completeness to selection function}
\label{sec:internal_completness}
While in this paper we mainly investigate the internal completeness of \rvs\ sample with respect to \all\ sample and specifically the \all\ sample with \magrp\ measurement (because we require \col\ colour), evidence shows \citep{2018ascl.soft11018R} that we can use this internal completeness and approximate the external completeness (i.e. selection function) from it.

\begin{figure}
	\includegraphics[width=\linewidth]{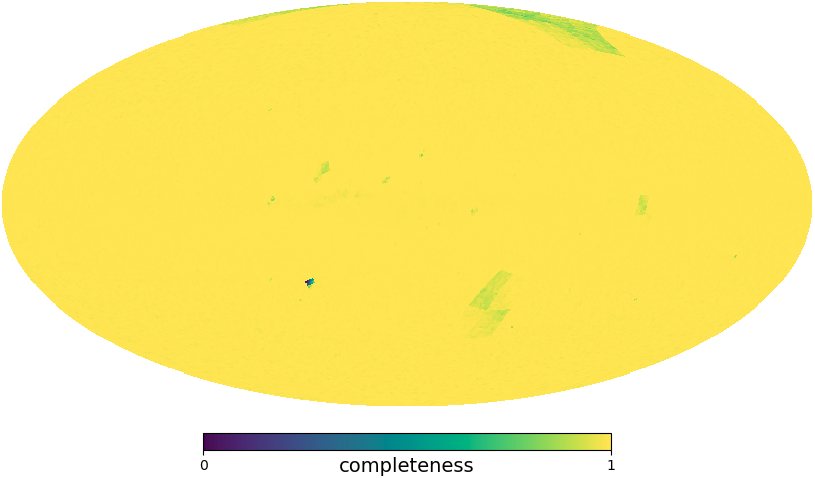}
	\caption{Internal completeness of the \all\ sample with \magrp\ measurement (of which \rvs\ sample is a subset of) versus the \all\ sample projected onto the sky (as a Mollweide projection as in Fig.\,\ref{fig:magcolorcut} at HEALpix level 6) for G$\,<15$\,mag. 
	}
	\label{fig:internal_completeness_grp}
\end{figure}

First we need to look at the internal completeness of \all\ sources that have a \magrp\ measurement with respect to \all\ sources that at least have a G measurement for G\,$<15$\,mag. As we see in Figure\,\ref{fig:internal_completeness_grp}, the completeness over the sky is virtually complete except for small patches with lower completeness; these are mainly towards the Galactic north pole plus a tiny patch with missing colour information.  

Further on we need to verify that \all\ sample is externally complete in the magnitude range brighter than 15\textsuperscript{th} G magnitude. \citet{2018ascl.soft11018R} did this via a crossmatch to 2MASS \citep{2006AJ....131.1163S} sources. Figures for this can be inspected in tutorial [2]\footnote{\url{https://github.com/jan-rybizki/gdr2_completeness/blob/master/tutorials/\%5B2\%5DCompleteness\%20tutorial_gdr2_light.ipynb}}. While they find external completeness of \all\ to be virtually complete between \(8 < G < 15\) mag, Gaia seems to be losing sources in the Galactic disc and bulge for sources with G\,$>15$ mag. Their external completeness assessment relies on the assumption that Gaia and 2MASS do independent measurements of the true sources within their respective G magnitude bins per HEALpix (they marginalise over colour). Despite their crude assumptions and large magnitude bins their findings are indicative of the \all\ sample being close to the true completeness for sources with G$\,<15\,$mag. Similarly we expect no bias from spurious sources in \all\ because for \magg\,$<15$\,mag these have negligible contribution, cf. Appendix C of \citet{2018A&A...616A...2L}.

For the bright end, \citet{2020MNRAS.tmp.2380B} have shown that for sources with G\,$>3$ mag the detection probability (and therefore completeness) is high.

We provide the \all\ sources with \magrp\ measurement internal completeness as a function of G and HEALpix of level 6 and integrate it into our RVS internal completeness function so that it should reflect the overall selection function.

\section{Accessing \rvs\ completeness function}
\label{sec:selection_function}
An example query illustrating how to download the data necessary to produce the \rvs\ selection function is given in Appendix\,\ref{sec:queries}.
User-specific quality cuts should be included in the query.

The internal completeness function created for easy usage as a python function has been generated in the following way: 
\begin{itemize}
    \item fixed bin size of 0.2\,mag in \magrvs\ covering a range from 2.9 until 14.1\,mag.
    \item fixed bin size of 0.1\ in \col, covering 0.05 to 1.75\,mag.
    \item HEALpix level 6 baseline but with degradation down to level 0 until at least 5 \all\ sources are available in that bin. If level 0 HEALpix has less than 5 sources then we use the whole sky also including bins with less than 5 sources.
    \item In each level 6 HEALpix the number of \rvs- and \all-sources are saved according to the above scheme as well as the respective HEALpix level from which the numbers were taken.
\end{itemize}

The completeness function as well as the upper and lower 1-$\sigma$ percentile were generated according to the following scheme:
    \begin{itemize}
        \item Whenever a bin had no \all\ or no \rvs\ entries the completeness function was set to zero as well as the upper and lower percentile.
        \item in the rest of the cases the number of sources in the \rvs\ bin were divided by the number of sources of the \all\ bin which yields the completeness function.
        \item for the upper and lower percentile we assumed a Poisson distribution with expected value of the number of sources in the \rvs\ bin. We took the value at the 16th and 84th percentile of this distribution and divided by the number of sources in the \all\ bin. We forced the value for upper to be less or equal the number of \all.
        
        
    \end{itemize}

\begin{figure*}
\includegraphics[width=\linewidth]{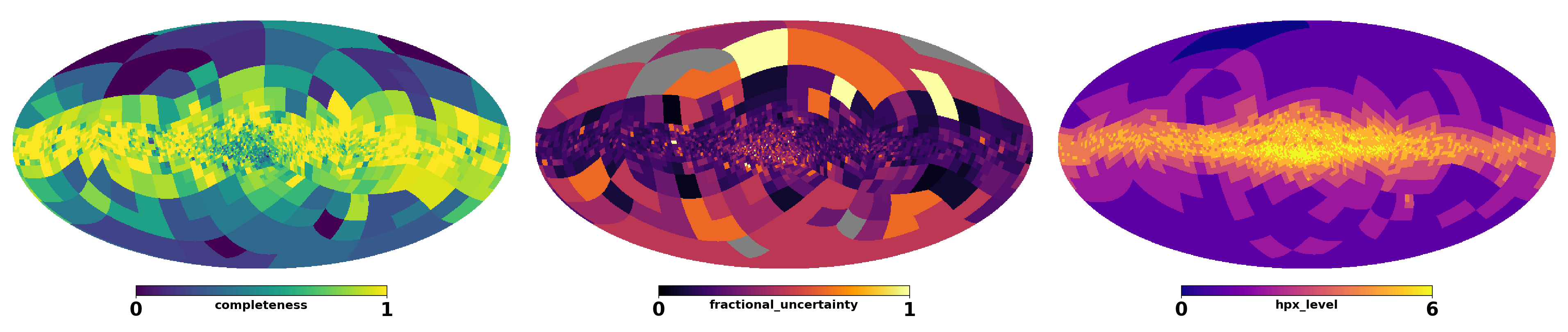}
 	\caption{Example selection function (all-sky output without RP to G completeness correction) for \magrvs\,=\,10.84\,mag and \col\,=\,1.2\,mag. From left to right, the completeness, fractional uncertainty ((upper-lower)/(2*completeness)) and the HEALpix level over which the star counts were averaged to calculate the completeness. The grey areas in the middle panel come from a zero division where the completeness function is zero.}
	\label{fig:selection_function}
\end{figure*}

As a result, the completeness function is smoothed on the sky wherever source densities were too low.An illustration of this can be seen in the left panel of Figure\,\ref{fig:selection_function}, where the whole sky selection function is depicted for \magrvs\,=\,10.84\,mag and \col\,=1.2\,mag. In the middle panel we show the corresponding fractional uncertainty calculated as ((upper-lower)/2)/completeness. The right panel shows the respective HEALpix level, over which the star counts in the CMD bin have been averaged. The \rvs\ completeness function can be queried (l, b, mag, col) and returns the above quantities and also the number of \all\ and \rvs\ sources in that respective bin (adding all of those yields total star counts), but also the sky area smoothed value (which is actually used to calculate the selection function). This is returned to a precision of 49152 HEALpixes, 56 magnitude bins and 17 colour bins and can be accessed via the \texttt{gdr2\_completeness} package \citep{2018ascl.soft11018R}. Tutorial\,5 illustrates its usage and shows some visualisations. We also provide the colour transformations $\left(\mathrm{G},\mathrm{G}_\mathrm{RP}\right)\rightarrow$\,\magrvs\ as well as the internal completeness of the \all\ sample with \magrp\ measurement vs the \all\ sample. Interactive web visualisations of the completeness maps over the CMD and the sky are available here\footnote{\url{https://www2.mpia-hd.mpg.de/homes/rybizki/index.html\#publ}}  

\section{Summary}
\label{sec:summary}
The \rvs\ sample is an important subset of the Gaia DR2 catalogue. We have characterized the internal completeness of the \rvs\ sample with respect to the \all\ sample in the three dimensions sky position (represented by HEALpix), \magrvs\ magnitude and \col\ colour. In the magnitude range $3<$\,\magrvs\,[mag]\,$<14$ evidence shows \citep{2018ascl.soft11018R} that this should be close to the external completeness (i.e. selection function). We show that the internal completeness is well characterised in the native \magrvs\ band which can be approximated from \magrp\ and \col. Imprints in the \rvs\ sample internal completeness arise from the IGSL3 input catalogues (mainly SDSS and GSC23) and its estimated external \magrvs\ on which the magnitude limit of 12\textsuperscript{th} magnitude has been applied. Completeness is lowered in high density areas mainly owing to blended RVS spectral windows resulting in non-rectangular windows which have not been processed in GDR2. Another important effect is the spectral template range between 3550 and 6900\,K that could be used to determine radial velocity measurements. Together with dust reddening in the Galactic plane this leads to the counter-intuitive effect that for red sources the completeness is best in the Galactic plane. A second-order effect arising from the scanning law and the requirement of at least 2 transits with radial velocity measurement is also visible.

We show that the \rvs\ sample can have partially different properties over the CMD than the corresponding \all\ sample, e.g. in parallax but also in proper motion (not depicted here, though), which might bias kinematic selections. We similarly show that \rvs\ sources usually have better quality indicators such as \texttt{ruwe} and \texttt{phot\_bp\_rp\_excess\_noise} compared to the \all\ sample which prohibits easy completeness corrections. We apply the completeness function to a mock stellar catalogue, GeDR3mock, and explore the impact of the parallax difference on the spatial extent of colour subsets of the \rvs\ sample. 

We provide a completeness function in \texttt{python} that delivers the internal completeness as a function of HEALpix, magnitude and colour together with quality and uncertainty indicators, together with the functionality to generate the external completeness, i.e. the selection function, by taking into account missing \magrp measurements in Gaia DR2. The necessary data are included but can also be generated for individual use-cases by adapting our example ADQL query from Appendix\,\ref{sec:queries}.

In Gaia DR3 (which is anticipated to be published end of 2021 at time of writing), a much improved \rvs\ sample will be provided. It will be much deeper with a magnitude limit of internal \magrvs$\,=14$\,mag and with the dependency on external catalogs removed. Also the treatment of blended spectra will be included and the effective temperature range for which radial velocities will be determined might increase. This will greatly simplify future selection function determinations and should allow for more sophisticated modelling, e.g. including the nearby source contamination (as projected onto the sky) and taking into account the different scanning angles.

\section*{Acknowledgements}
We thank David Katz for insightful discussions on the RVS selection function. We also thank the Gaia group at MPIA for valuable feedback and Berry Holl for a very helpful email exchange on the Gaia scanning law. 
We thank the anonymous referee (also over a second iteration) for their thorough inspection and helpful comments. JR will not travel anywhere by aeroplane for the purpose of promoting this paper.

This work has made use of data from the European Space Agency (ESA) mission Gaia, processed by the Gaia Data Processing and Analysis Consortium (DPAC). Funding for the DPAC has been provided by national institutions, in particular the institutions participating in the Gaia Multilateral Agreement. 

This research or product makes use of public auxiliary data provided by ESA/Gaia/DPAC as obtained from the publicly accessible ESA Gaia SFTP.

This work was funded by the DLR (German space agency) via grant 50\,QG\,1403.

WJC studentship provided by University of Hertfordshire and Istituto Nazionale di Astrofisica, Osservatorio Astrofisico di Torino with computing infrastructure provided by STFC grant ST/R000905/1.

Software: \texttt{topcat} \citep{2005ASPC..347...29T}, \texttt{HEALpix} \citep{2005ApJ...622..759G},
\texttt{astropy} \citep{2018AJ....156..123A}, \texttt{corner} \citep{2016JOSS....1...24F}.

\subsection*{Data availability} 
The data underlying this article are available in the article and in its online supplementary material.


\bibliographystyle{mnras}
\bibliography{RVS_selection} 

\begin{thebibliography}{}
\makeatletter
\relax
\def\mn@urlcharsother{\let\do\@makeother \do\$\do\&\do\#\do\^\do\_\do\%\do\~}
\def\mn@doi{\begingroup\mn@urlcharsother \@ifnextchar [ {\mn@doi@}
  {\mn@doi@[]}}
\def\mn@doi@[#1]#2{\def\@tempa{#1}\ifx\@tempa\@empty \href
  {http://dx.doi.org/#2} {doi:#2}\else \href {http://dx.doi.org/#2} {#1}\fi
  \endgroup}
\def\mn@eprint#1#2{\mn@eprint@#1:#2::\@nil}
\def\mn@eprint@arXiv#1{\href {http://arxiv.org/abs/#1} {{\tt arXiv:#1}}}
\def\mn@eprint@dblp#1{\href {http://dblp.uni-trier.de/rec/bibtex/#1.xml}
  {dblp:#1}}
\def\mn@eprint@#1:#2:#3:#4\@nil{\def\@tempa {#1}\def\@tempb {#2}\def\@tempc
  {#3}\ifx \@tempc \@empty \let \@tempc \@tempb \let \@tempb \@tempa \fi \ifx
  \@tempb \@empty \def\@tempb {arXiv}\fi \@ifundefined
  {mn@eprint@\@tempb}{\@tempb:\@tempc}{\expandafter \expandafter \csname
  mn@eprint@\@tempb\endcsname \expandafter{\@tempc}}}

\bibitem[\protect\citeauthoryear{{Andrae} et~al.,}{{Andrae}
  et~al.}{2018}]{2018A&A...616A...8A}
{Andrae} R.,  et~al., 2018, \mn@doi [\aap] {10.1051/0004-6361/201732516}, \href
  {https://ui.adsabs.harvard.edu/abs/2018A&A...616A...8A} {616, A8}

\bibitem[\protect\citeauthoryear{{Arenou} et~al.,}{{Arenou}
  et~al.}{2018}]{2018A&A...616A..17A}
{Arenou} F.,  et~al., 2018, \mn@doi [\aap] {10.1051/0004-6361/201833234}, \href
  {https://ui.adsabs.harvard.edu/abs/2018A&A...616A..17A} {616, A17}

\bibitem[\protect\citeauthoryear{{Astropy Collaboration} et~al.,}{{Astropy
  Collaboration} et~al.}{2018}]{2018AJ....156..123A}
{Astropy Collaboration} et~al., 2018, \mn@doi [\aj] {10.3847/1538-3881/aabc4f},
  \href {https://ui.adsabs.harvard.edu/abs/2018AJ....156..123A} {156, 123}

\bibitem[\protect\citeauthoryear{{Boubert} \& {Everall}}{{Boubert} \&
  {Everall}}{2020}]{2020MNRAS.tmp.2380B}
{Boubert} D.,  {Everall} A.,  2020, \mn@doi [\mnras] {10.1093/mnras/staa2305},
  \href {https://ui.adsabs.harvard.edu/abs/2020MNRAS.tmp.2380B} {}

\bibitem[\protect\citeauthoryear{{Boubert}, {Everall}  \& {Holl}}{{Boubert}
  et~al.}{2020}]{2020MNRAS.497.1826B}
{Boubert} D.,  {Everall} A.,   {Holl} B.,  2020, \mn@doi [\mnras]
  {10.1093/mnras/staa2050}, \href
  {https://ui.adsabs.harvard.edu/abs/2020MNRAS.497.1826B} {497, 1826}

\bibitem[\protect\citeauthoryear{{Brandeker} \& {Cataldi}}{{Brandeker} \&
  {Cataldi}}{2019}]{2019A&A...621A..86B}
{Brandeker} A.,  {Cataldi} G.,  2019, \mn@doi [\aap]
  {10.1051/0004-6361/201834321}, \href
  {https://ui.adsabs.harvard.edu/abs/2019A&A...621A..86B} {621, A86}

\bibitem[\protect\citeauthoryear{{Cropper} et~al.,}{{Cropper}
  et~al.}{2018}]{2018A&A...616A...5C}
{Cropper} M.,  et~al., 2018, \mn@doi [\aap] {10.1051/0004-6361/201832763},
  \href {https://ui.adsabs.harvard.edu/abs/2018A&A...616A...5C} {616, A5}

\bibitem[\protect\citeauthoryear{{Evans} et~al.,}{{Evans}
  et~al.}{2018}]{2018A&A...616A...4E}
{Evans} D.~W.,  et~al., 2018, \mn@doi [\aap] {10.1051/0004-6361/201832756},
  \href {https://ui.adsabs.harvard.edu/abs/2018A&A...616A...4E} {616, A4}

\bibitem[\protect\citeauthoryear{{Foreman-Mackey}}{{Foreman-Mackey}}{2016}]{2016JOSS....1...24F}
{Foreman-Mackey} D.,  2016, \mn@doi [The Journal of Open Source Software]
  {10.21105/joss.00024}, \href
  {https://ui.adsabs.harvard.edu/abs/2016JOSS....1...24F} {1, 24}

\bibitem[\protect\citeauthoryear{{Gaia Collaboration} et~al.,}{{Gaia
  Collaboration} et~al.}{2016}]{2016A&A...595A...2G}
{Gaia Collaboration} et~al., 2016, \mn@doi [\aap]
  {10.1051/0004-6361/201629512}, \href
  {https://ui.adsabs.harvard.edu/abs/2016A&A...595A...2G} {595, A2}

\bibitem[\protect\citeauthoryear{{Gaia Collaboration} et~al.,}{{Gaia
  Collaboration} et~al.}{2018}]{2018A&A...616A...1G}
{Gaia Collaboration} et~al., 2018, \mn@doi [\aap]
  {10.1051/0004-6361/201833051}, \href
  {https://ui.adsabs.harvard.edu/abs/2018A&A...616A...1G} {616, A1}

\bibitem[\protect\citeauthoryear{{G{\'o}rski}, {Hivon}, {Banday}, {Wandelt},
  {Hansen}, {Reinecke}  \& {Bartelmann}}{{G{\'o}rski}
  et~al.}{2005}]{2005ApJ...622..759G}
{G{\'o}rski} K.~M.,  {Hivon} E.,  {Banday} A.~J.,  {Wandelt} B.~D.,  {Hansen}
  F.~K.,  {Reinecke} M.,   {Bartelmann} M.,  2005, \mn@doi [\apj]
  {10.1086/427976}, \href {http://adsabs.harvard.edu/abs/2005ApJ...622..759G}
  {622, 759}

\bibitem[\protect\citeauthoryear{{H{\o}g} et~al.,}{{H{\o}g}
  et~al.}{2000}]{2000A&A...355L..27H}
{H{\o}g} E.,  et~al., 2000, \aap, \href
  {https://ui.adsabs.harvard.edu/abs/2000A&A...355L..27H} {355, L27}

\bibitem[\protect\citeauthoryear{{Katz} et~al.,}{{Katz}
  et~al.}{2019}]{2019A&A...622A.205K}
{Katz} D.,  et~al., 2019, \mn@doi [\aap] {10.1051/0004-6361/201833273}, \href
  {https://ui.adsabs.harvard.edu/abs/2019A&A...622A.205K} {622, A205}

\bibitem[\protect\citeauthoryear{{Lasker} et~al.,}{{Lasker}
  et~al.}{2008}]{2008AJ....136..735L}
{Lasker} B.~M.,  et~al., 2008, \mn@doi [\aj] {10.1088/0004-6256/136/2/735},
  \href {https://ui.adsabs.harvard.edu/abs/2008AJ....136..735L} {136, 735}

\bibitem[\protect\citeauthoryear{{Lindegren} et~al.,}{{Lindegren}
  et~al.}{2018}]{2018A&A...616A...2L}
{Lindegren} L.,  et~al., 2018, \mn@doi [\aap] {10.1051/0004-6361/201832727},
  \href {https://ui.adsabs.harvard.edu/abs/2018A&A...616A...2L} {616, A2}

\bibitem[\protect\citeauthoryear{{Prusti}}{{Prusti}}{2012}]{2012AN....333..453P}
{Prusti} T.,  2012, \mn@doi [Astronomische Nachrichten]
  {10.1002/asna.201211688}, \href
  {https://ui.adsabs.harvard.edu/abs/2012AN....333..453P} {333, 453}

\bibitem[\protect\citeauthoryear{{Riello} et~al.,}{{Riello}
  et~al.}{2018}]{2018A&A...616A...3R}
{Riello} M.,  et~al., 2018, \mn@doi [\aap] {10.1051/0004-6361/201832712}, \href
  {https://ui.adsabs.harvard.edu/abs/2018A&A...616A...3R} {616, A3}

\bibitem[\protect\citeauthoryear{{Rybizki} \& {Drimmel}}{{Rybizki} \&
  {Drimmel}}{2018}]{2018ascl.soft11018R}
{Rybizki} J.,  {Drimmel} R.,  2018, {gdr2\_completeness (v2.0): GaiaDR2 data
  retrieval and manipulation} (\mn@eprint {ascl} {1811.018})

\bibitem[\protect\citeauthoryear{{Rybizki} et~al.,}{{Rybizki}
  et~al.}{2020}]{2020PASP..132g4501R}
{Rybizki} J.,  et~al., 2020, \mn@doi [\pasp] {10.1088/1538-3873/ab8cb0}, \href
  {https://ui.adsabs.harvard.edu/abs/2020PASP..132g4501R} {132, 074501}

\bibitem[\protect\citeauthoryear{{Sartoretti} et~al.,}{{Sartoretti}
  et~al.}{2018}]{2018A&A...616A...6S}
{Sartoretti} P.,  et~al., 2018, \mn@doi [\aap] {10.1051/0004-6361/201832836},
  \href {https://ui.adsabs.harvard.edu/abs/2018A&A...616A...6S} {616, A6}

\bibitem[\protect\citeauthoryear{{Skrutskie} et~al.,}{{Skrutskie}
  et~al.}{2006}]{2006AJ....131.1163S}
{Skrutskie} M.~F.,  et~al., 2006, \mn@doi [\aj] {10.1086/498708}, \href
  {https://ui.adsabs.harvard.edu/abs/2006AJ....131.1163S} {131, 1163}

\bibitem[\protect\citeauthoryear{{Smart} \& {Nicastro}}{{Smart} \&
  {Nicastro}}{2014}]{2014A&A...570A..87S}
{Smart} R.~L.,  {Nicastro} L.,  2014, \mn@doi [\aap]
  {10.1051/0004-6361/201424241}, \href
  {https://ui.adsabs.harvard.edu/abs/2014A&A...570A..87S} {570, A87}

\bibitem[\protect\citeauthoryear{{Strauss} et~al.,}{{Strauss}
  et~al.}{2002}]{2002AJ....124.1810S}
{Strauss} M.~A.,  et~al., 2002, \mn@doi [\aj] {10.1086/342343}, \href
  {https://ui.adsabs.harvard.edu/abs/2002AJ....124.1810S} {124, 1810}

\bibitem[\protect\citeauthoryear{{Taylor}}{{Taylor}}{2005}]{2005ASPC..347...29T}
{Taylor} M.~B.,  2005, in {Shopbell} P.,  {Britton} M.,   {Ebert} R.,  eds,
  Astronomical Society of the Pacific Conference Series Vol. 347, Astronomical
  Data Analysis Software and Systems XIV. p.~29

\bibitem[\protect\citeauthoryear{{de Bruijne}, {Allen}, {Azaz},
  {Krone-Martins}, {Prod'homme}  \& {Hestroffer}}{{de Bruijne}
  et~al.}{2015}]{2015A&A...576A..74D}
{de Bruijne} J.~H.~J.,  {Allen} M.,  {Azaz} S.,  {Krone-Martins} A.,
  {Prod'homme} T.,   {Hestroffer} D.,  2015, \mn@doi [\aap]
  {10.1051/0004-6361/201424018}, \href
  {https://ui.adsabs.harvard.edu/abs/2015A&A...576A..74D} {576, A74}

\makeatother
\end{thebibliography}




\appendix

\section{Querying the Selection function}
\label{sec:queries}
This query uses Common Table Expressions, which will be part of the upcoming ADQL 2.1 standard, which at the time of writing among the Gaia-carrying VO data centers are only available on GAVO's TAP service\footnote{\url{http://www.g-vo.org/}}. 

\begin{lstlisting}
 WITH with_rvs AS ( -- This is a subquery on which we will perform further queries below
    SELECT radial_velocity,	source_id/140737488355328 AS hpx, phot_g_mean_mag-phot_rp_mean_mag AS grp,
    -- Creates HEALpix of level 6 and abbreviates the G-GRP colour
  	  phot_rp_mean_mag+0.042319-0.65124*(phot_g_mean_mag - phot_rp_mean_mag)+ 1.0215 * POWER(phot_g_mean_mag - phot_rp_mean_mag,2) - 1.3947 * POWER(phot_g_mean_mag - phot_rp_mean_mag,3) + 0.53768 * POWER(phot_g_mean_mag - phot_rp_mean_mag,4) AS phot_rvs -- GRVS approximation
    FROM gaia.dr2light -- This is the GAVO Gaia DR2 table, which contains only essential columns
    WHERE phot_g_mean_mag-phot_rp_mean_mag<1.4 AND phot_g_mean_mag-phot_rp_mean_mag>0.05
  UNION ALL -- Above the GRVS approximation for blue sources and below for red sources, here we join them
    SELECT radial_velocity,	source_id/140737488355328 AS hpx, phot_g_mean_mag-phot_rp_mean_mag AS grp,
  	  phot_rp_mean_mag+132.32-377.28*(phot_g_mean_mag - phot_rp_mean_mag)+ 402.32 * POWER(phot_g_mean_mag - phot_rp_mean_mag,2) - 190.97 * POWER(phot_g_mean_mag - phot_rp_mean_mag,3) + 34.026 * POWER(phot_g_mean_mag - phot_rp_mean_mag,4) AS phot_rvs
    FROM gaia.dr2light
    WHERE phot_g_mean_mag-phot_rp_mean_mag>=1.4	AND phot_g_mean_mag-phot_rp_mean_mag<1.75),
  hasrvs AS ( -- New subquery on which we will perform yet another query below
	  SELECT hpx, ROUND(phot_rvs/2.0,1)*2.0 AS mag, ROUND(grp,1) AS colour, COUNT(*) AS n_withrvs
	  -- Setting the bin size in GRVS and colour and getting the source count per bin
	  FROM with_rvs -- For all sources with radial velocity measurement in the respective bin
	  WHERE radial_velocity IS NOT NULL	AND phot_rvs > 2.9 AND phot_rvs < 14.1
	  GROUP BY hpx, mag, colour),
  allobjs AS (select hpx, ROUND(phot_rvs/2.0,1)*2.0 AS mag, ROUND(grp,1) AS colour, COUNT(*) AS n_all
	  FROM with_rvs -- For all sources in Gaia DR2 in the respective bin
	WHERE phot_rvs > 2.9 AND phot_rvs < 14.1 -- Within the magnitude range
	  GROUP BY hpx, mag, colour)
 SELECT hpx, mag, colour, n_all, n_withrvs
  FROM hasrvs RIGHT OUTER JOIN allobjs USING (hpx, mag, colour)
  -- Collecting the necessary entries from above for subsequent processing in python
\end{lstlisting}

G$_\mathrm{RVS}$ is approximated using Equation 2 \& 3 from \citet{2018A&A...616A...1G}.

The autocorrelation query for the close pairs is as follows:
\begin{lstlisting}
 WITH with_rvs AS ( -- Overall a similar construct to the above query    
	SELECT radial_velocity, source_id, ra, dec, phot_g_mean_mag, phot_rp_mean_mag,
		phot_rp_mean_mag+0.042319-0.65124*(phot_g_mean_mag - phot_rp_mean_mag)
			+ 1.0215 * POWER(phot_g_mean_mag - phot_rp_mean_mag,2) - 1.3947 
			* POWER(phot_g_mean_mag - phot_rp_mean_mag,3) + 0.53768 
			* POWER(phot_g_mean_mag - phot_rp_mean_mag,4) AS phot_rvs
	FROM gaia.dr2light
	WHERE phot_g_mean_mag-phot_rp_mean_mag<1.4 AND phot_g_mean_mag-phot_rp_mean_mag>0.05
UNION ALL
	SELECT radial_velocity, source_id, ra, dec, phot_g_mean_mag, phot_rp_mean_mag,
		phot_rp_mean_mag+132.32-377.28*(phot_g_mean_mag - phot_rp_mean_mag)
			+ 402.32 * POWER(phot_g_mean_mag - phot_rp_mean_mag,2) - 190.97 
			* POWER(phot_g_mean_mag - phot_rp_mean_mag,3) + 34.026 
			* POWER(phot_g_mean_mag - phot_rp_mean_mag,4) AS phot_rvs
	FROM gaia.dr2light
	WHERE phot_g_mean_mag-phot_rp_mean_mag>=1.4 AND phot_g_mean_mag-phot_rp_mean_mag<1.75),
rvobject AS (
SELECT source_id, ra, dec, phot_rvs, phot_g_mean_mag, phot_rp_mean_mag, radial_velocity
FROM with_rvs
WHERE radial_velocity IS NOT NULL)
SELECT a.source_id AS sid1, b.source_id AS sid2, a.phot_rvs AS rvsmag1, a.ra AS ra1, a.dec AS dec1, a.phot_g_mean_mag AS gmag1, a.phot_rp_mean_mag AS rpmag1, a.radial_velocity AS rv1, b.phot_g_mean_mag AS gmag2, b.phot_rp_mean_mag AS rpmag2, b.radial_velocity AS rv2, DISTANCE(a.ra, a.dec, b.ra, b.dec)*3600 AS dist
FROM rvobject AS a -- Here we only request pairs with an angular separation of less than 20 as.
     JOIN gaia.dr2light AS b
     ON (a.source_id!=b.source_id AND DISTANCE(b.ra, b.dec, a.ra, a.dec)<20/3600.)
     -- If adding any WHERE statement here the source_id index gets lost and the query takes long
\end{lstlisting}
The result of this query (which was subdivided in different HEALpix in order to be able to retrieve all sources) can be downloaded from here\footnote{\url{https://keeper.mpdl.mpg.de/d/f2b841c75d7a42f6aad9/}} in a 5, 10 and 20\arcsec\ version. It has also been cleaned from double entries and an \magrvs\ magnitude has been calculated for the second source.

\section{Increased densities from second Field of View}
\label{sec:FOV2}
Since the two telescopes share the focal plane, the sources competing for CCD window allocation are more than just the sources at a single position in the sky. We neglect this effect in our work, because the viewing angle of 106.5 degrees is sufficiently large and the position of the second telescope always changes for a specific position of the first telescope because of the scanning law. Therefore, plus the fact that most fractions of the sky contain low density areas, we assume that only a small fraction of transits will have the situation where both fields are in crowded regions.

To make an approximate but quantitative assessment of the increase in source density per transit coming from the second telescope, we use the public Gaia scanning law\footnote{\url{https://www.cosmos.esa.int/web/gaia/scanning-law-pointings}} \citep{2016A&A...595A...2G}, without accounting for the breaks (e.g. due to lost telemetry or the telescope going into safe mode) and estimate: for each field of view for telescope 1 (FoV1) on the sky, how many more sources there are from FoV2 (on average). We take every 5th data point in the scanning law file (which has a time resolution of 10 seconds), such that the individual pointings are 50\arcmin\ apart (50 seconds), which corresponds approximately to the distance of two neighbouring HEALpix at level 6. Level 6 HEALpix also have roughly the size of the field of view (FoV) of one telescope. Instead of an exact solution, each pointing is moved to the nearest HEALpix of level 6. We assume that per transit in each FoV Gaia observes the same amount of sources, which is simply taken from the HEALpix level 6 source count of \rvs\ sample (as displayed in Figure\,\ref{fig:igsl_rvs}).

\begin{figure}
	\includegraphics[width=\linewidth]{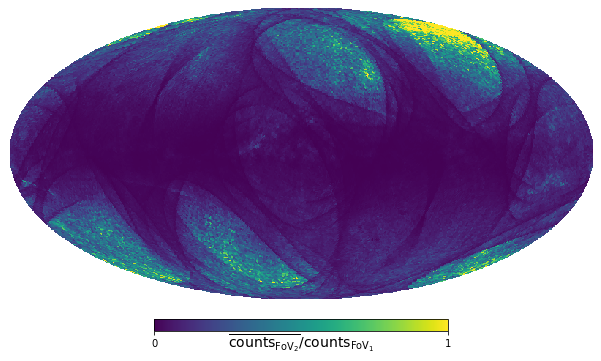}
	\caption{Mollweide projection of the mean per transit fractional increase in source density for \rvs\ sample owing to the second telescope in Galactic coordinates. The Galactic Center is in the middle, with longitude increasing to the left at HEALpix level 6.}
	\label{fig:scanlaw}
\end{figure}

As can be seen in Figure\,\ref{fig:scanlaw} the resulting mean per transit fractional increase for the \rvs\ sample (average counts in FoV2 divided by counts in FoV1) is very low in high density regions, cf. right panel Figure\,\ref{fig:igsl_rvs}. Therefore neglecting the sources from the second telescope in high density regions seems to be a valid assumption. For the \all\ sample the situation is similar and will even improve in future data releases with longer observational baselines. The shown plot and others can be recreated using tutorial\,6 in \citet{2018ascl.soft11018R}.

\bsp	
\label{lastpage}
\end{document}